\documentclass{PoS}

\title{High luminosity fixed-target experiment at the LHC}

\ShortTitle{High luminosity fixed-target experiment at the LHC}

 \author{\speaker{C.~Hadjidakis}$^{1}$, S.J.~Brodsky$^{2}$, G.~Cavoto$^{3}$, C.~Da Silva$^{4}$, F.~Donato$^{5}$, M.G.~Echevarria$^{6}$, E.G.~Ferreiro$^{7}$, I.~H\v{r}ivn\'{a}\v{c}ov\'{a}$^{1}$, D.~Kikola$^{8}$, A.~Klein$^{4}$, A.~Kurepin$^{9}$, A.~Kusina$^{10}$, J.P.~Lansberg$^{1}$, C.~Lorc\'e$^{11}$, F.~Lyonnet$^{12}$, Y.~Makdisi$^{13}$, L.~Massacrier$^{1}$, S.~Porteboeuf$^{14}$, C.~Quintans$^{15}$, A.~Rakotozafindrabe$^{16}$, P.~Robbe$^{17}$, W.~Scandale$^{18}$, I.~Schienbein$^{19}$, J.~Seixas$^{15,20}$, H.S.~Shao$^{21}$, A.~Signori$^{22}$, N.~Topilskaya$^{9}$, B.~Trzeciak$^{23}$, A.~Uras$^{24}$, J.~Wagner$^{25}$, N.~Yamanaka$^{1}$, Z.~Yang$^{26}$, A.~Zelenski$^{27}$\\
 {\tiny 
\llap{$^1$}IPNO, CNRS/IN2P3, Univ. Paris-Sud, Universit\'e Paris-Saclay,  Orsay, France\\
\llap{$^2$}
          SLAC National\,Accelerator\,Laboratory, Stanford University, Menlo Park, USA\\
\llap{$^{3}$}
	"Sapienza" Universit\`a di Roma, Dipartimento di Fisica \& INFN, Sez. di Roma, P.le A. Moro 2, 00185 Roma, Italy\\
\llap{$^4$}
          LANL, P-25, Los Alamos National Laboratory, Los Alamos, NM 87545, USA\\
\llap{$^{5}$}
	Turin University, Department of Physics, and INFN, Sezione of Turin, Turin, Italy\\
\llap{$^6$}
         INFN Sez. Pavia, Via Bassi 6, 27100 Pavia, Italy\\
\llap{$^7$} 
         Dept. de F{\'\i}sica de Part{\'\i}culas, USC, Santiago de Compostella, Spain\\   
\llap{$^8$}           
          Faculty of Physics, Warsaw University of Technology,  Warsaw, Poland\\ 
\llap{$^{9}$}      
          Institute for Nuclear Research, Russian Academy of Sciences, Moscow, Russia\\                      
\llap{$^{10}$} 
        Institute of Nuclear Physics Polish Academy of Sciences, PL-31342 Krakow, Poland\\               
\llap{$^{11}$}   
         CPhT, Ecole Polytechnique, CNRS, Universit\'e Paris-Saclay,  Palaiseau, France\\
\llap{$^{12}$}    
         Southern Methodist University, Dallas, TX 75275, USA\\   
\llap{$^{13}$}
	Brookhaven National Laboratory, Collider Accelerator Department\\     
\llap{$^{14}$}
	Universit\'e Clermont Auvergne, CNRS/IN2P3, LPC, F-63000 Clermont-Ferrand, France\\
\llap{$^{15}$}     
        LIP and IST, Lisbon, Portugal\\   
\llap{$^{16}$}
	IRFU/DPhN, CEA Saclay, 91191 Gif-sur-Yvette Cedex, France\\        
\llap{$^{17}$}         
         LAL, Univ. Paris-Sud, CNRS/IN2P3, Universit\'e Paris-Saclay,  Orsay, France\\   
\llap{$^{18}$} 
	CERN, European Organization for Nuclear Research, 1211 Geneva 23, Switzerland\\
\llap{$^{19}$}           
         LPSC, Universit\'e Grenoble-Alpes, CNRS/IN2P3, 38026 Grenoble, France\\    
\llap{$^{20}$} 
	Dep. Fisica, Instituto Superior Tecnico, Av. Rovisco Pais 1, 1049-001 Lisboa, Portugal\\
\llap{$^{21}$} 
         Laboratoire de Physique Th\'eorique et Hautes \'Energies (LPTHE), UMR 7589, Sorbonne Universit\'e et CNRS, 4 place Jussieu, 75252 Paris Cedex 05, France\\
\llap{$^{22}$} 
      Theory Center, Thomas Jefferson National Accelerator Facility, 12000 Jefferson Avenue, Newport News, VA 23606, USA\\      
\llap{$^{23}$} 
       Institute for Subatomic Physics, Utrecht University, Utrecht, The Netherlands\\           
\llap{$^{24}$} 
      IPNL, Universit\'e Claude Bernard Lyon-I, CNRS/IN2P3, Villeurbanne, France\\       
\llap{$^{25}$}        
       National Centre for Nuclear Research (NCBJ), Hoza 69, 00-681, Warsaw, Poland\\       
\llap{$^26$}          
          CHEP, Department of Engineering Physics, Tsinghua University, Beijing, China\\
\llap{$^{27}$}
	Brookhaven National Laboratory, Collider Accelerator Department\\
}
}

\abstract{By extracting the beam with a bent crystal or by using an internal gas target, the multi-TeV proton and lead LHC beams allow one to perform the most energetic fixed-target experiments ever and to study $pp$, $p$d and $p$A collisions at $\sqrt{s_{NN}}=115$~GeV and Pb$p$ and PbA collisions at $\sqrt{s_{NN}}=72$~GeV with high precision and modern detection techniques. Such studies would address open questions in the domain of the nucleon and nucleus partonic structure at high-$x$, quark-gluon plasma and, by using longitudinally or transversally polarised targets, spin physics. In this paper, we will review the technical solutions to obtain a high-luminosity fixed-target experiment at the LHC and will discuss their possible implementations with the ALICE and LHCb detectors.}

\FullConference{International Conference on Hard and Electromagnetic Probes of High-Energy Nuclear Collisions\\
		30 September - 5 October 2018\\
		Aix-Les-Bains, Savoie, France}

\begin{document}

\section{Physics motivations}

Fixed-target experiments offer many advantages having the versatility of polarised and nuclear targets and allowing to 
reach high-luminosity with dense and long target. 
The AFTER@LHC project aims at demonstrating the physics opportunities and the technical implementations 
of a high-luminosity fixed-target experiment using the LHC beams. The 7 TeV proton and 2.76 A TeV lead beams 
allow one to reach a center-of-mass energy per nucleon pair of 
$\sqrt{s_{NN}}$~=~115~GeV and $\sqrt{s_{NN}}$~=~72~GeV with a center-of-mass rapidity boost of 4.8 and 4.2 units, respectively. These energies correspond to an energy domain between SPS and nominal RHIC energies. The large rapidity boost implies that the mid- to forward rapidity region in the center of mass frame ($y_{cms} \geq 0$) lies within 1 degree in the laboratory frame and that the backward rapidity region ($y_{cms} \leq 0$) is easily accessible by using standard experimental techniques 
or existing LHC experiments such as ALICE or LHCb. Thus the fixed-target mode at high-luminosity presents unique opportunities to access the very backward rapidity domain and therefore the high-$x$ frontier, where $x$ is the momentum fraction of the parton struck in the target nucleon or nucleus. 

The physics motivations of a fixed-target experiment at the LHC are three-fold: advance our understanding of the high-$x$ gluon, antiquark and heavy-quark content in the nucleon and nucleus, unravel the spin structure of the nucleon, and study the quark-gluon plasma (QGP) created in heavy-ion collisions towards large rapidity~\cite{Brodsky:2012vg,Massacrier:2015qba,Trzeciak:2017csa,Kikola:2017hnp,Hadjidakis:2018ifr} (see also~\cite{webpage} for more details). 
In the high-$x$ programme, one could extract parton distribution functions (PDFs), poorly known in this kinematic regime, by using Drell-Yan measurements to probe the light quark, and open heavy-flavour measurements to probe the gluon and charm content of the nucleon. 
In a nuclear target, one could reveal the EMC effect which is far from understood and determine whether such an effect exists also for the gluon case. With a transversally polarised target, one could access information on orbital motion of partons bound into hadrons by measuring the Sivers effect. 
This effect can be probed for light quark with Drell-Yan, or for gluon with open heavy-flavour production. Finally in Pb-A collisions at $\sqrt{s_{NN}}$~=~72~GeV, the medium created is expected to have a low baryon chemical potential and a temperature approximately 1.5 times higher than the critical temperature of the phase transition between a hadron gas and a QGP~\cite{Satz}. At such temperature, the excited states of $\Upsilon$ are expected to melt into the QGP~\cite{Mocsy}. In addition, measurements of particle yields and their anisotropies as a function of rapidity and system size would allow one to scan the phase-transition region. In the following, we will concentrate on the technical implementations and projects under investigation in ALICE and LHCb. For the physics topics, the reader can refer to the recent review~\cite{Hadjidakis:2018ifr} as well as to other contributions from the Hard Probes 2018 conference~\cite{afterTalk}.

\section{Possible technical implementations}

Several techniques are promising to obtain a fixed-target experiment at the LHC. 
LHCb has pioneered the use of gaseous fixed-target with the SMOG system~\cite{smog,smog2,smog3}, originally designed for luminosity calibration. In that case, the gas density is low since the gas is not confined to a specific region and there is no dedicated pumping system. Also only noble gases have been used so far and for limited running time periods. Higher density gas targets are possible with the gas-jet system or by using a storage cell. With those target setups, H, D and $^3$He gases could be injected as well as polarised gases. 
Solid targets can also provide high luminosity fixed-target experiment. The beam halo can interact directly with the target inserted inside the beam pipe. Another more promising solution, that has the advantage to be more parasitic to the main beams, is to deviate the beam halo on an internal solid target by using a bent crystal. In that case, the proton and lead fluxes are expected to be approximately $5.10^8$ p/s and $2.10^5$ Pb/s and would allow one to obtain large luminosities with a target of a few millimeter thickness. With this technique, two solutions can be envisioned: either the beam is extracted and a new beam line is created, or the beam is used on a target located in an existing cavern (beam splitting). The first solution requires however civil engineering and could be achieved only over a longer timescale. The second solution is possible provided that the deflected beam halo is then absorbed upstream of the detector. 
The implementations of the gas and solid target solutions at the LHC were investigated in~\cite{Hadjidakis:2018ifr} and are currently discussed in the Physics Beyond Collider working groups at CERN, in particular with the aim at evaluating the effect on the LHC beams~\cite{ipac}. As we will see in the following, these possible technical implementations would allow one to set up a fixed-target programme with the current LHC detectors, ALICE and LHCb. 

\section{ALICE and LHCb in a fixed-target running mode}

\begin{figure}[hbt!]
\centering
\includegraphics[width=0.58\textwidth]{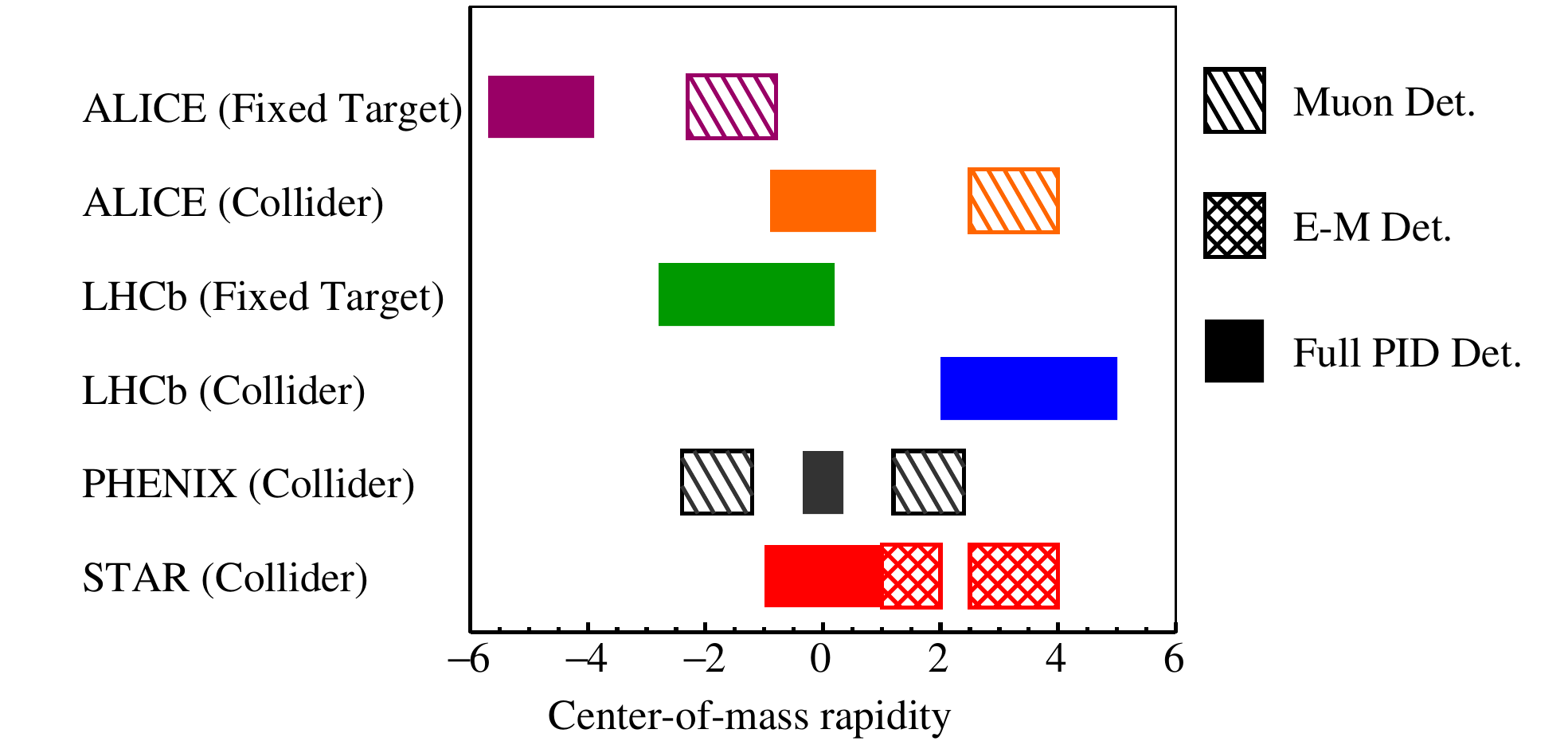} 
\includegraphics[width=0.38\textwidth]{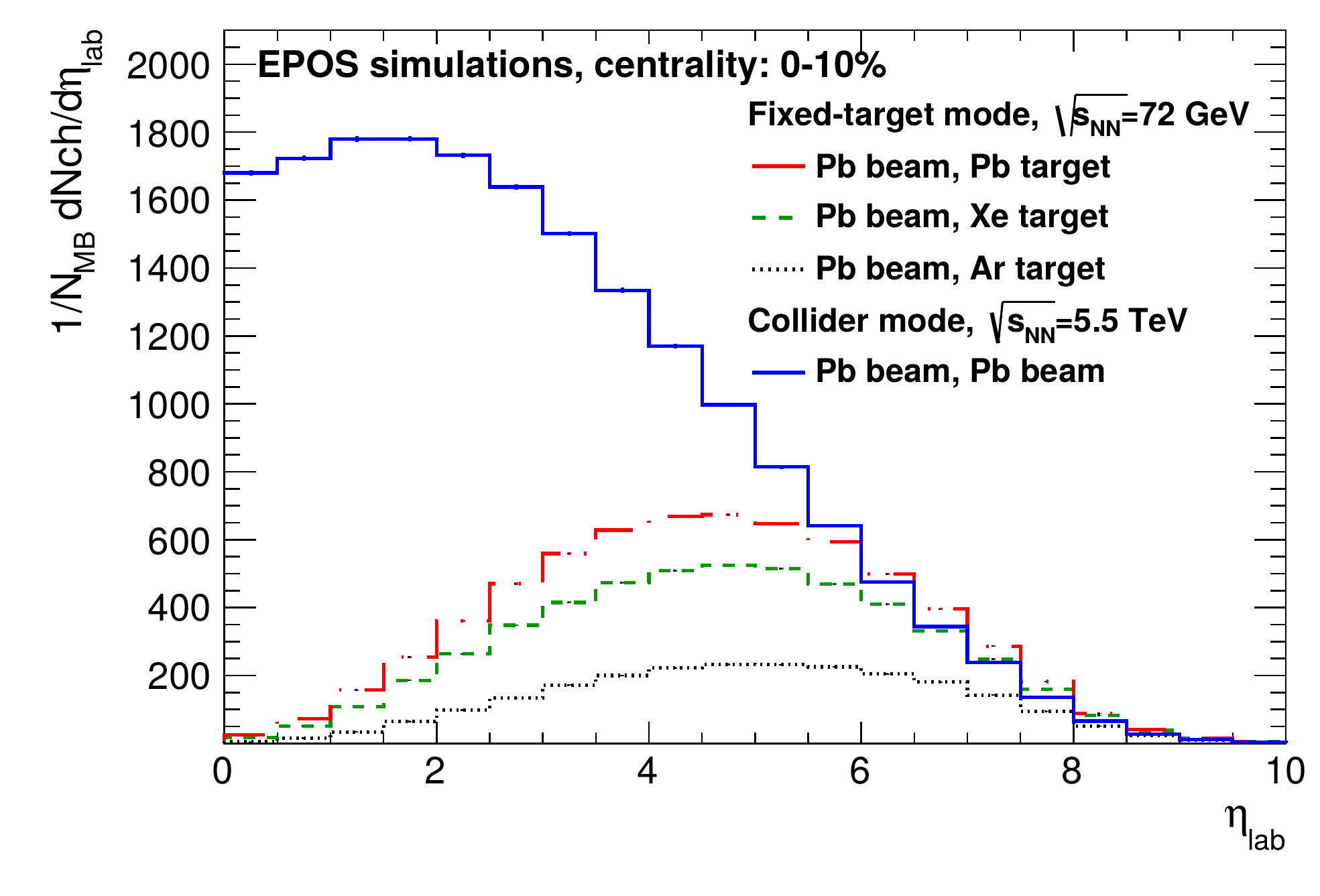}
\caption{Left panel: Comparison of the rapidity coverage of the ALICE and LHCb detectors at LHC, and the STAR and PHENIX detectors at RHIC. For ALICE and LHCb, the acceptance is shown in collider and fixed-target modes with a target position at the nominal Interaction Point for a 7 TeV proton beam. Right panel: Averaged charged-particle multiplicity as a function of the pseudorapidity in the laboratory frame for various heavy-ion systems.}
\label{fig:detector}
\end{figure}

Most of the physics cases outlined in~\cite{Hadjidakis:2018ifr} can be covered with the ALICE and LHCb detectors if they are coupled with a fixed target setup. 
The rapidity boost implies the particles are mostly produced in the forward direction. The left panel of Fig.~\ref{fig:detector} shows the rapidity acceptance of the ALICE and LHCb detectors in the collider and fixed-target modes where one can see, in the latter case, the effect of the rapidity boost. The rapidity coverages are also compared to the ones of the STAR and PHENIX detector at RHIC. One of the main advantage of the fixed-target mode at the LHC is that particles can be easily detected at large angles corresponding to very large values of negative $y_{cms}$. In a fixed-target mode, LHCb detectors give access to a wide rapidity range starting from mid-rapidity in the center of mass. In ALICE, the central barrel (full square) covers the very backward center-of-mass rapidity and the muon spectrometer (dashed square) covers a rapidity interval towards mid-rapidity.     
Table~\ref{table_lumi} shows the yearly integrated luminosities that could be delivered to ALICE and LHCb for various target types with the proton and lead beams. We considered as limitations to the obtained luminosities either the technical solution or the detector data-taking-rates capabilities. More details can be found in~\cite{Hadjidakis:2018ifr}. In ALICE, 1 MHz and 50 kHz of minimum bias data-taking rates were considered as maxima for the proton and lead beams, respectively. In LHCb, a maximum of 40 MHz was considered for the proton beam while a rate of approximately 200 kHz in the Pb-Xe case is expected to be sustainable by the experiment. The resulting rates are large and comparable to the ones expected in the collider mode for Run3 and 4. 
\begin{table}[!hbt]
\small
\centering
\begin{tabular}{c | c | c | c | c | c }
\multicolumn{2}{c|}{Target } & p beam & Pb beam & p beam & Pb beam\\
Technique & Type &  $\int \mathcal{L}_{\rm ALICE}$ & $\int \mathcal{L}_{\rm ALICE}$ & $\int \mathcal{L}_{\rm LHCb}$& $\int \mathcal{L}_{\rm LHCb}$\\
\hline
	    & $\rm{H}^\uparrow$ & 43 pb$^{-1}$ &   0.56 nb$^{-1}$ & 43 pb$^{-1}$ &   0.56 nb$^{-1}$ \\ 
Gas jet & $\rm{H}_2$ &  0.26 fb$^{-1}$ &   28 nb$^{-1}$ & 10 fb$^{-1}$ &   118 nb$^{-1}$\\ 
            & $\rm{Xe}$ &  7.7 pb$^{-1}$ &   8.1 nb$^{-1}$ & 0.31 fb$^{-1}$ &   23 nb$^{-1}$\\ 

\hline
	 & $\rm{H}^\uparrow$ &  0.26 fb$^{-1}$ &   28 nb$^{-1}$ & 9.2 fb$^{-1}$ &   118 nb$^{-1}$ \\ 
Storage cell & $\rm{H}_2$ &  0.26 fb$^{-1}$ &   28 nb$^{-1}$ & 10 fb$^{-1}$ &   118 nb$^{-1}$\\ 
            & $\rm{Xe}$ &  7.7 pb$^{-1}$ &   8.1 nb$^{-1}$  & 0.31 fb$^{-1}$ &   30 nb$^{-1}$  \\ 
\hline
	   & $\rm{C}$ (658 $\mu$m) &  37 pb$^{-1}$ &   $-$  & $-$ &   $-$ \\
Bent crystal   & $\rm{C}$ (5 mm) &  $-$ &   5.6 nb$^{-1}$  & 280 pb$^{-1}$ &   5.6 nb$^{-1}$\\
 and solid target     &  $\rm{W}$ (184 $\mu$m) &  5.9 pb$^{-1}$ &   $-$  & $-$ &   $-$ \\ 
      &  $\rm{W}$ (5 mm) &  $-$ &   3.1 nb$^{-1}$  & 160 pb$^{-1}$ &   3.1 nb$^{-1}$ \\ 

\hline
\end{tabular}
\caption{Summary of the achievable integrated yearly luminosities for some technical implementations and targets with the ALICE and LHCb detectors in the fixed target mode, accounting for the data-taking-rate capabilities (see text). The integrated luminosity corresponds to a LHC year with time duration of ${\rm t}_{\rm p}~=~10^7$~s and ${\rm t}_{\rm Pb}~=~10^6$~s for the proton and lead beams, respectively.}
\label{table_lumi}
\end{table}
In the case of the lead beam with a nuclear target, the detector should be able to cope with high occupancy in order to reconstruct high multiplicity events. The right panel of Fig.~\ref{fig:detector} compares the pseudorapidity dependence of the charged particle multiplicity of different systems in the fixed-target mode at $\sqrt{s_{NN}}$~=~72~GeV with the one of the Pb-Pb collider mode at $\sqrt{s_{NN}}$~=~5.5~TeV, in the case of the 10\% more central collisions. These distributions are obtained from EPOS~\cite{Werner:2005jf,Pierog:2013ria} simulations. In the fixed-target mode, the average charged multiplicity does not exceed the one of the LHC collider mode. The ALICE detectors can reconstruct events at such high-multiplicity. In LHCb, the event reconstruction is so far limited to the 50\% less central events in Pb-Pb collisions at $\sqrt{s_{NN}}=5$~TeV. 

Three projects are under investigation in LHCb: an internal solid target coupled to a crystal with a second crystal close to the target for EDM/MDM experiments~\cite{Neri:2019JanuaryPBC}, a storage cell attached to the VELO (SMOG2) that will provide two order of magnitude higher gas pressure with respect to SMOG~\cite{Graziani:2019JanuaryPBC}, and a polarised storage cell target for spin physics~\cite{Graziani:2019JanuaryPBC}. The installation of the SMOG2 system is foreseen in LS2. In ALICE, the internal solid target coupled to a bent crystal is investigated with a target location inside the L3 solenoid magnet few meters upstream of IP2~\cite{Kikola:2019JanuaryPBC}.  

\section{Conclusion}
Many physics opportunities are offered by a high-luminosity fixed-target experiment at the LHC with detectors covering a wide rapidity range. The achievable luminosities would permit decisive measurements both on quark and gluon sensitive probes such as Drell-Yan, open heavy-flavour and quarkonium production. With transversally polarised targets, it will be possible to study the quark and gluon Sivers functions by measuring the single transverse spin asymmetries for Drell-Yan, open and hidden heavy-flavour production. The lead beam interaction on a nuclear target would allow one to study the quark-gluon plasma in an energy range between the SPS and the top RHIC energy over a wide rapidity range. Various technical implementations are possible and some of them are promising to obtain high luminosities, such as the internal gas target or the internal solid target coupled with the beam halo deflection by a bent crystal. In LHCb, three projects are under investigation (crystal and internal solid target, unpolarised and polarised gas targets). In particular, SMOG2 will be installed in LS2 and will provide unpolarised gas target at high density. In ALICE, the crystal and internal solid target solution is currently being investigated.  
\\

{\bf Acknowledgement} {\small This research was supported by the French P2IO Excellence Laboratory, the French CNRS via the grants FCPP-Quarkonium4AFTER \& D\'efi Inphyniti-Th\'eorie LHC France, the P2I department of the Paris-Saclay University, the COPIN-IN2P3 Agreement, 
by the RFBR/CNRS grant 18-52-15007 (Russia and France) and by the U.S. Department of Energy, Contract DE-AC02-76SF00515. AS acknowledges support from U.S. Department of Energy contract DE-AC05-06OR23177, under which Jefferson Science Associates, LLC, manages and operates Jefferson Lab.}

\end{document}